\newcommand\apjcls{1}
\newcommand\aastexcls{2}
\newcommand\othercls{3}

\newcommand\papercls{\aastexcls}
\documentclass[tighten, times, twocolumn]{aastex62}  

\if\papercls \apjcls
\usepackage{apjfonts}
\else\if\papercls \othercls
\usepackage{epsfig}
\usepackage{margin}
\usepackage{times}
\fi\fi
\usepackage{ifthen}
\usepackage{natbib}
\usepackage{amssymb, amsmath}
\usepackage{appendix}
\usepackage{etoolbox}
\usepackage[T1]{fontenc}
\usepackage{paralist}
\usepackage{upgreek}

\if\papercls \apjcls
\newcommand\aas{\ref@jnl{AAS Meeting Abstracts}}
\newcommand\dps{\ref@jnl{AAS/DPS Meeting Abstracts}}
\newcommand\maps{\ref@jnl{MAPS}}
\else\if\papercls \othercls
\usepackage{astjnlabbrev-jh}
\fi\fi

\if\papercls \aastexcls
\hypersetup{citecolor=blue, 
            linkcolor=blue, 
            menucolor=blue, 
            urlcolor=blue}  
\else
\usepackage[
bookmarks=true,           
bookmarksnumbered=true,   
colorlinks=true,          
citecolor=blue,           
linkcolor=blue,           
menucolor=blue,           
urlcolor=blue,            
linkbordercolor={0 0 1},  
pdfborder={0 0 1},
frenchlinks=true]{hyperref}
\fi

\if\papercls \othercls

\else

\fi

\providecommand{\adsurl}[1]{\href{#1}{ADS}}

\makeatletter
\patchcmd{\NAT@citex}
  {\@citea\NAT@hyper@{%
     \NAT@nmfmt{\NAT@nm}%
     \hyper@natlinkbreak{\NAT@aysep\NAT@spacechar}{\@citeb\@extra@b@citeb}%
     \NAT@date}}
  {\@citea\NAT@nmfmt{\NAT@nm}%
   \NAT@aysep\NAT@spacechar\NAT@hyper@{\NAT@date}}{}{}

\patchcmd{\NAT@citex}
  {\@citea\NAT@hyper@{%
     \NAT@nmfmt{\NAT@nm}%
     \hyper@natlinkbreak{\NAT@spacechar\NAT@@open\if*#1*\else#1\NAT@spacechar\fi}%
       {\@citeb\@extra@b@citeb}%
     \NAT@date}}
  {\@citea\NAT@nmfmt{\NAT@nm}%
   \NAT@spacechar\NAT@@open\if*#1*\else#1\NAT@spacechar\fi\NAT@hyper@{\NAT@date}}
  {}{}
\makeatother

\makeatletter
\DeclareRobustCommand{\lowcase}[1]{\@lowcase#1\@nil}
\def\@lowcase#1\@nil{\if\relax#1\relax\else\MakeLowercase{#1}\fi}
\makeatother

\DeclareSymbolFont{UPM}{U}{eur}{m}{n}
\DeclareMathSymbol{\umu}{0}{UPM}{"16}
\let\oldumu=\umu
\renewcommand\umu{\ifmmode\oldumu\else\math{\oldumu}\fi}

\if\papercls \othercls

\else

\fi

\let\oldsim=\sim
\renewcommand\sim{\ifmmode\oldsim\else\math{\oldsim}\fi}
\let\oldpm=\pm
\renewcommand\pm{\ifmmode\oldpm\else\math{\oldpm}\fi}
\newcommand\by{\ifmmode\times\else\math{\times}\fi}

\newbox{\wdbox}
\renewcommand\c{\setbox\wdbox=\hbox{,}\hspace{\wd\wdbox}}
\renewcommand\i{\setbox\wdbox=\hbox{i}\hspace{\wd\wdbox}}

\newcount\timect
\newcount\hourct
\newcount\minct
\newcommand\now{\timect=\time \divide\timect by 60
         \hourct=\timect \multiply\hourct by 60
         \minct=\time \advance\minct by -\hourct
         \number\timect:\ifnum \minct < 10 0\fi\number\minct}

\catcode`@=11

\newcommand\comment[1]{}

\newcommand\commenton{\catcode`\%=14}

\renewcommand\math[1]{$#1$}
\newcommand\mathshifton{\catcode`\$=3}

\let\atab=&
\newcommand\atabon{\catcode`\&=4}

\let\oldmsp=\sp
\let\oldmsb=\sb
\def\sp#1{\ifmmode
           \oldmsp{#1}%
         \else\strut\raise.85ex\hbox{\scriptsize #1}\fi}
\def\sb#1{\ifmmode
           \oldmsb{#1}%
         \else\strut\raise-.54ex\hbox{\scriptsize #1}\fi}
\newbox\@sp
\newbox\@sb
\def\sbp#1#2{\ifmmode%
           \oldmsb{#1}\oldmsp{#2}%
         \else
           \setbox\@sb=\hbox{\sb{#1}}%
           \setbox\@sp=\hbox{\sp{#2}}%
           \rlap{\copy\@sb}\copy\@sp
           \ifdim \wd\@sb >\wd\@sp
             \hskip -\wd\@sp \hskip \wd\@sb
           \fi
        \fi}
\def\msp#1{\ifmmode
           \oldmsp{#1}
         \else \math{\oldmsp{#1}}\fi}
\def\msb#1{\ifmmode
           \oldmsb{#1}
         \else \math{\oldmsb{#1}}\fi}

\def\supon{\catcode`\^=7}

\def\subon{\catcode`\_=8}

\def\supsubon{\supon \subon}

\newcommand\actcharon{\catcode`\~=13}

\newcommand\paramon{\catcode`\#=6}

\comment{And now to turn us totally on and off...}

\newcommand\reservedcharson{ \commenton  \mathshifton  \atabon  \supsubon 
                             \actcharon  \paramon}

\catcode`@=12
\reservedcharson

\if\papercls \apjcls

\else

\fi

\newcommand\chisq{\ifmmode{\chi\sp{2}}\else\math{\chi\sp{2}}\fi}
\newcommand\redchisq{\ifmmode{ \chi\sp{2}\sb{\rm red}}
                    \else\math{\chi\sp{2}\sb{\rm red}}\fi}
\newcommand\Teq{\ifmmode{T\sb{\rm eq}}\else$T$\sb{eq}\fi}
\newcommand\mjup{\ifmmode{M\sb{\rm Jup}}\else$M$\sb{Jup}\fi}
\newcommand\rjup{\ifmmode{R\sb{\rm Jup}}\else$R$\sb{Jup}\fi}
\newcommand\msun{\ifmmode{M\sb{\odot}}\else$M\sb{\odot}$\fi}
\newcommand\rsun{\ifmmode{R\sb{\odot}}\else$R\sb{\odot}$\fi}
\newcommand\mearth{\ifmmode{M\sb{\oplus}}\else$M\sb{\oplus}$\fi}
\newcommand\rearth{\ifmmode{R\sb{\oplus}}\else$R\sb{\oplus}$\fi}

\newcommand{\norm}[1]{\left\lVert#1\right\rVert}
\newcommand{\Rgas}{R_{\rm gas}}
\newcommand{\muG}{\upmu {\rm G}}
\shorttitle{Analytic Magnetic Field}
\shortauthors{Bino \& Basu 2020}

\begin{document}

\title{Fitting an Analytic Magnetic Field to a Prestellar Core}

\author{Gianfranco Bino}
\affiliation{Department of Applied Mathematics, University of Western Ontario,  London, ON, N6A 5B7, Canada.}

\author{Shantanu Basu}
\affiliation{Department of Physics \& Astronomy, University of Western Ontario, London, ON, N6A 3K7, Canada.}
\affiliation{Institute for Earth \& Space Exploration, University of Western Ontario, London, Ontario N6A 5B7, Canada.}


\email{gbino@uwo.ca}
\email{basu@uwo.ca}


\begin{abstract}
We deploy and demonstrate the capabilities of the magnetic field model developed by \citet{ewe13} by fitting observed polarimetry data of the prestellar core FeSt 1--457. The analytic hourglass magnetic field function derived directly from Maxwell's equations yields a central-to-surface magnetic field strength ratio in the equatorial plane, as well as magnetic field directions with relative magnitudes throughout the core. This fit emerges from a comparison of a single plane of the model with the polarization map that results from the integrated properties of the magnetic field and dust throughout the core. Importantly, our fit is independent of any assumed density profile of the core. We check the robustness of the fit by using the POLARIS code to create synthetic polarization maps that result from the integrated scattering and emission properties of the dust grains and their radiative transfer, employing an observationally-motivated density profile. We find that the synthetic polarization maps obtained from the model also provides a good fit to the observed polarimetry. Our model fits the striking feature of significant curvature of magnetic field lines in the outer part of FeSt 1--457. Combined with independent column density estimates, we infer that the core of size $\Rgas$ has a mildly supercritical mass-to-flux ratio and may have formed through dynamical motions starting from a significantly larger radius $R$. A breakdown of flux-freezing through neutral-ion slip (ambipolar diffusion) could be responsible for effecting such a transition from a large-scale magnetic field structure to a more compact gas structure.
\end{abstract}


\section{Introduction}
\label{introduction}

Hourglass magnetic fields are a direct result of gravitational contraction and the flux-freezing property of plasma that is governed by ideal magnetohydrodynamics (MHD). An hourglass pattern may also arise from magnetic induction in a contracting cloud in which nonideal MHD processes are at work. \citet{bas09} showed that the degree of curvature of an hourglass pattern would depend on the amount of nonideal MHD effects at work. 
These and other numerical MHD models \citep[e.g.,][]{kud07,kud11} yield hourglass shapes, but which not available in a closed analytic form. 

Polarimetry is a powerful tool to elucidate magnetic field morphology and reveal hourglass or other systematic magnetic field patterns. This has been known since
\citet{dav51} proposed that elongated paramagnetic dust grains would generally spin about their
minor axis and that this axis would come into alignment with the ambient magnetic field through paramagnetic relaxation. Although theories of the detailed mechanism have evolved \citep[see][]{and15}, and are complicated by the lack of detailed knowledge of grain composition, the outcome is 
qualitatively similar in all theories. This (at least partial) alignment of grains leads to dust emission with electric field perpendicular to the local magnetic field direction, or absorption of background starlight (dichroic extinction) that leaves a net polarization of the electric field that is parallel to the local magnetic field.  
The hourglass morphology has been observed through polarimetry in many dense star-forming regions, including the protostellar envelope NGC 1333 IRAS 4A \citep{gir06} and the massive star-forming regions OMC-1 \citep{sch98} and G31.41+0.31 \citep{gir09}. These systems contain embedded stellar objects, which are thought to aid the alignment mechanism \citep{and15}. Recently, \citet{kan17} measured an hourglass polarization pattern for the first time in a starless core, FeSt 1--457. They used near-infrared polarimetry of the dichroic extinction, utlizing the location of the core in the Pipe Nebula region and in the foreground of the Galactic center region with a rich set of thousands of background stars. While polarization measurements can reveal a magnetic field pattern, the polarization fraction cannot be used to directly measure the field strength due to the many uncertainties in dust grain properties and the alignment mechanism and efficiency. An indirect means of field strength determination from polarimetry is through the Davis-Chandrasekhar-Fermi (DCF) method \citep{davis51,cha53} that uses a model of Alfv\'enic fluctuations and measurements of the density, velocity dispersion, and magnetic field direction dispersion to estimate a magnetic field strength. \citet{kan17} used this method to determine a magnetic field strength in FeSt 1--457 and found that the mass-to-flux ratio is mildly supercritical, but subject to order unity uncertainty due to the unknown inclination angle of the magnetic field. A more direct measure of the 
(line-of-sight component) magnetic field strength in gas clouds can be done through the Zeeman effect \citep{cru12}, but has proven to be quite difficult due to observational sensitivity and resolution. 
Improvements in the sensitivity and resolution of polarized emission measurements at far-infrared and submillimeter wavelengths are leading to significant advances in the ability to measure magnetic field morphology.  The \textit{Planck} satellite measurements of submillimeter polarization of Galactic dust emission \citep{pla16} reveals an ordered polarization pattern on large scales that also merges smoothly with the smaller-scale near-infrared polarization observations of molecular clouds. The patterns imply an ordered magnetic field that is consistent with sub-Alfv\'enic turbulence and mildly subcritical mass-to-flux ratio inferred using the DCF method \citep{pla16}.  

With an increasing set of maps of polarization direction \citep[e.g.,][]{ste13,mau18,soa18,red19,sad19} of 
 dense star-forming regions, it is important to develop techniques to infer the magnetic field properties and mass-to-flux ratios from the available data.
 Other properties like velocity dispersion \citep{pin10} are also relevant. For example, \citet{aud19} used the spatial variation of velocity dispersion to estimate the magnetic field strength and field dispersion inside dense cores. When an hourglass pattern is observed in the polarization map, there is in principle information about the degree of magnetic field strength enhancement from the background value to that of the innermost resolved region. \citet{mye18,mye20} have developed models of flux-frozen magnetic fields that condense from a uniform background medium into a spheroidal configuration under the assumption that the magnetic field strength $B$ scales with the density $\rho$ as $B \propto \rho^{2/3}$ at all locations. Although this is not a self-consistent mapping from the background state allowed by a continuous deformation of magnetic field and plasma, as calculated by \cite{mou76a,mou76b}, it does give an analytic form that is amenable to fitting observational data. \cite{kan6} have used the model of \cite{mye18} to fit the polarization segments of FeSt 1--457 and found that the best fit to the field lines required that the flux-frozen contraction would have started at a relatively high density of $\sim 5000\, {\rm cm}^{-3}$, quite a bit higher than the
 $\sim 300\, {\rm cm}^{-3}$ interclump density in the Pipe Nebula.

A common method in modeling the hourglass structures has been the fitting of a two dimensional slice of a mathematical model to the observed morphology of the polarization segments.
One approach is to fit the field with a set of nested parabolas \citep{gir06, kan17}, which is a purely mathematical matching.
Other models that a priori assume flux-freezing \citep{aud19b,mye18,mye20} will necessarily infer a central magnetic field enhancement that is related (through flux freezing) to the observed density or column density enhancement determined by other means, e.g., the intensity of dust emission.
In this paper, we utilize the analytic hourglass magnetic field derived by \citet{ewe13}. This model assumes an axisymmetric geometry and is a self-consistent solution of Maxwell's equations for a system with a current density that is confined near the $z=0$ plane in cylindrical geometry. Fitting the model to an inferred magnetic field direction map obtained through polarimetry yields the magnetic field components $B_r$ and $B_z$. However, such a comparison is based on matching a single plane of the model with the polarization map that represents an integral through the core of the magnetic field and dust distribution. The integration needs to account for the polarized emission and scattering properties of the dust grains and their radiative transfer. Therefore, we also use the POLARIS radiative transfer code \citep{rei16} to determine the emergent polarization directions, using a density profile of the form that has been used to fit the column density distribution of FeSt 1--457 \citep{kan05}. 
Furthermore, the magnetic field profile is not tied to any particular density profile, so the observed density profile and best fit magnetic field model do not have to match through any specified relation. Discrepancies in the layout of these distributions can imply the presence of nonideal MHD effects in the core formation and evolution.


\section{Methods}
\label{sec:methods}
\subsection{Magnetic Field Model Fitting}
The method to model the magnetic field using polarimetry data involves an approach that is slightly different from that done by \citet{ewe13}, who fit the model magnetic field components directly to simulation data. When dealing with polarimetry data we do not know the magnitude of each individual component of the magnetic field, but rather can estimate the ratio $B_r/B_z$. We begin with the expressions for the magnetic field components given in \citet{ewe13}:
\begin{align}
B_r (r,z) = \sum_{m=1}^{\infty} k_m \sqrt{\lambda_m} J_1(\sqrt{\lambda_m}r)\left[ \text{erfc} \left( \frac{\sqrt{\lambda_m }h}{2} - \frac{z}{h} \right) \right. \nonumber \\ \left. \times e^{-z\sqrt{\lambda_m}} - \text{erfc} \left( \frac{\sqrt{\lambda_m }h}{2}  + \frac{z}{h} \right)e^{z\sqrt{\lambda_m}} \right], \label{eq:m1}
\end{align}
\begin{align}
B_z(r,z)  = \sum_{m=1}^{\infty} k_m \sqrt{\lambda_m} J_0(\sqrt{\lambda_m}r)\left[ \text{erfc} \left( \frac{\sqrt{\lambda_m }h}{2} + \frac{z}{h} \right)\right. \nonumber \\ \left. \times e^{z\sqrt{\lambda_m}} + \text{erfc} \left( \frac{\sqrt{\lambda_m }h}{2}  - \frac{z}{h} \right)e^{-z\sqrt{\lambda_m}} \right] + B_0, \label{eq:m2}
\end{align} 
where $h$ is the scale height for the Gaussian $z$-distribution of electric current density, and $\lambda = \left( a_{\text{m},1} / R \right) ^2$, where $a_{\text{m},1}$ is the $m^{\text{th}}$ root of the Bessel function of the first kind. In this situation we can observe (measure) the relative angle that each polarization segment makes with the normal (the $z$-axis in our case) and use this offset angle to fit the ratio $B_r/B_z$, that is we minimize the quantity
\begin{align}
\Delta (r_i, z_i) = \sum_i \norm{\frac{B_r (r_i,z_i)}{B_z (r_i,z_i)} - \tan \phi_i}^2, \label{eq:res}
\end{align}
where the angles $\phi$ are data quantities and the ratio $B_r/B_z$ is taken from the model. We can express the magnetic field nondimensionally by defining the parameters
$$\tilde{r} = r/R, \:\:\:\:\:\:\:\: \tilde{z} = z/R, \:\:\:\:\:\:\:\: \eta = h/R,$$\vspace{-0.3cm}
$$\tilde{B}_r = B_r/B_0, \:\:\:\:\:\:\:\: \tilde{B}_z = B_z/B_0, \:\:\:\:\:\:\:\: \beta_m = k_m \sqrt{\lambda_m} / B_0,$$
where we normalize the magnetic field components to the background field and normalize the positional quantities $r$ and $z$ to the core radius $R$. These give rise to the nondimensionalized counterparts to Equations \eqref{eq:m1} and \eqref{eq:m2}:
\begin{align}
\tilde{B}_r = \sum_{m=1}^{\infty} \beta_m J_1(a_{\text{m},1} \tilde{r})\left[ \text{erfc} \left( \frac{a_{\text{m},1} \eta}{2} - \frac{\tilde{z}}{\eta} \right)e^{-a_{\text{m},1} \tilde{z}}  \right. \nonumber \\ \left. - \text{erfc} \left( \frac{a_{\text{m},1} \eta}{2} + \frac{\tilde{z}}{\eta} \right)e^{a_{\text{m},1} \tilde{z}} \right] ,
\label{dim:br}
\end{align}
\begin{align}
\tilde{B}_z = \sum_{m=1}^{\infty} \beta_m J_0(a_{\text{m},1} \tilde{r})\left[ \text{erfc} \left( \frac{a_{\text{m},1} \eta}{2} + \frac{\tilde{z}}{\eta} \right)e^{a_{\text{m},1} \tilde{z}}  \right. \nonumber \\ \left. + \text{erfc} \left( \frac{a_{\text{m},1} \eta}{2} - \frac{\tilde{z}}{\eta} \right)e^{-a_{\text{m},1} \tilde{z}} \right] + 1 .
\label{dim:bz}
\end{align}
This formalism gives freedom to the model user to select a value of $B_0$ and have each $\beta_m$ normalized to that value. To perform the optimization we use the Sequential Quadratic Programming algorithm, part of Matlab's \texttt{fmincon} function, to minimize Equation \eqref{eq:res}.
\subsection{POLARIS Model}
\label{sec:polarismodel}

We can use our  models to produce synthetic polarization maps that simulate the magnetic field as a three-dimensional integral through the entire core rather than taking a slice of a single plane as done in Figure \ref{fig:3}. To perform the simulations we used the three dimensional radiative transfer code POLARIS developed by \cite{rei16}. We include grain alignment in the simulated polarization maps and restrict the alignment to be perfectly perpendicular to the magnetic field. This allows one to directly infer the magnetic field morphology through the orientation of dust grains. We model the gas profile with a modified isothermal sphere:
\begin{equation}
\rho (r) = \frac{\rho_c}{1 + (r/a)^2}, 
\end{equation}
with central density $\rho_c = m_n\,n_c$, central number density $n_c = 3.5 \times 10^5$ cm$^{-3}$ \citep{kan05}, and neutral mass $m_n = 2.3\, m_{\rm H}$. The radial scale length $a$ is defined by $a^2 = A\,c_s^2/(2 \pi G \rho_c)$ where $c_s$ is the isothermal sound speed. We adopt $A=2$ so as to achieve a Bonnor-Ebert-like density profile with a gas mass $M_{\rm gas} \approx 3.6\,M_{\odot}$ within a radius $\Rgas \approx 0.1$ pc, in agreement with the fit of the gas density profile for FeSt 1--457 obtained by \cite{kan05}. For a detailed discussion of the relation between a modified isothermal sphere and a Bonnor-Ebert sphere, see \cite{dap09}.

The grain alignment by radiative torques requires an anisotropic radiation source \citep[see][]{and15}. Therefore, we implement a weakly radiating protostar of temperature $T = 300\, \text{K}$ and radius $R = 4R_{\odot}$ as the radiation source for our simulations, mimicking a possible observationally undetected first hydrostatic core \citep[see, e.g.,][]{eno2010}. The grain size distribution is taken to follow an MRN \citep{mat77} power-law distribution in which the number density of grains $n_{\rm g}$ satisfies $dn_{\rm g}/da \propto n_{\rm g}^{-3.5}$ from a minimum grain size $a_{\rm min}=0.005\, \upmu{\rm m}$ to a maximum grain size $a_{\rm max}=2\, \upmu{\rm m}$. We adopt a dust-to-gas mass ratio of 0.01. These and other properties are listed in Table \ref{tab:1}, where we also adopt the gas and dust temperatures from \cite{kan17}.
\begin{figure*}[h]
 \centering
 \includegraphics[width = 0.465\textwidth]{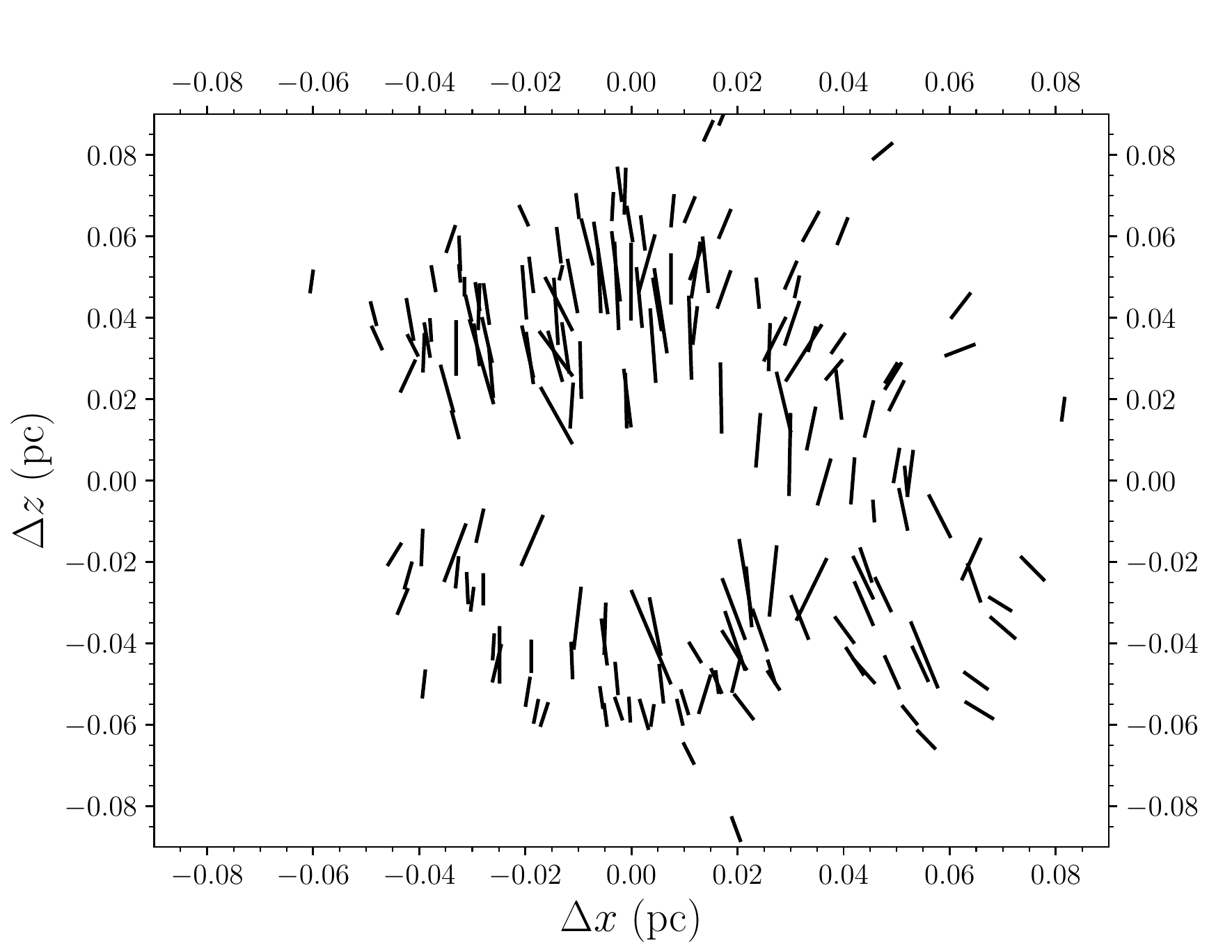}
  \caption{Polarimetry data for the core FeSt 1--457 \citep{kan17}. Each dark line corresponds to polarization in the near-infrared due to dichroic extinction of light from a background star. The direction of each polarization segment is taken to be parallel to the magnetic field. The lengths represent the degree of polarization, with the longest line corresponding to about 5\% polarization.} 
	\label{fig:pol}
 \end{figure*}
 \begin{figure*}
  \hspace{1cm}
\begin{minipage}{\textwidth}
\includegraphics[width = 0.9\textwidth]{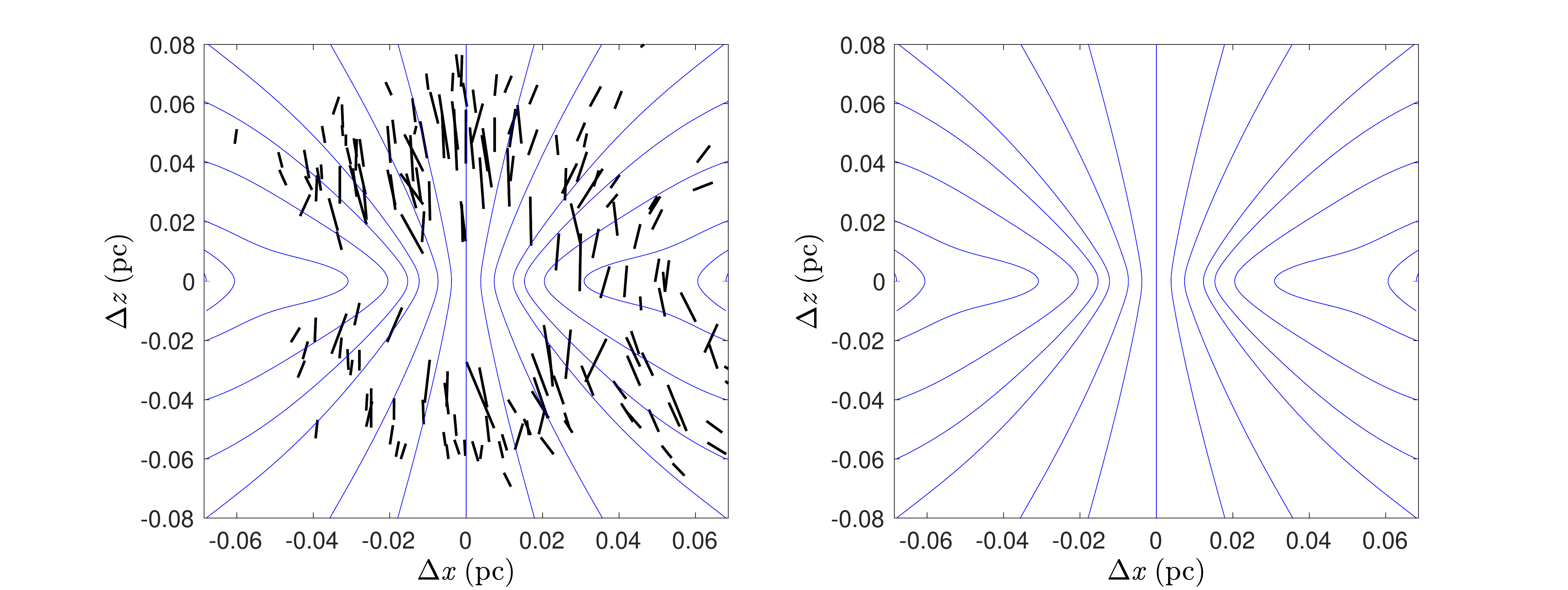}
\end{minipage}
\begin{minipage}{\textwidth}
\vspace{-0.45cm}
\hspace{1cm}
\includegraphics[width=0.9\textwidth]{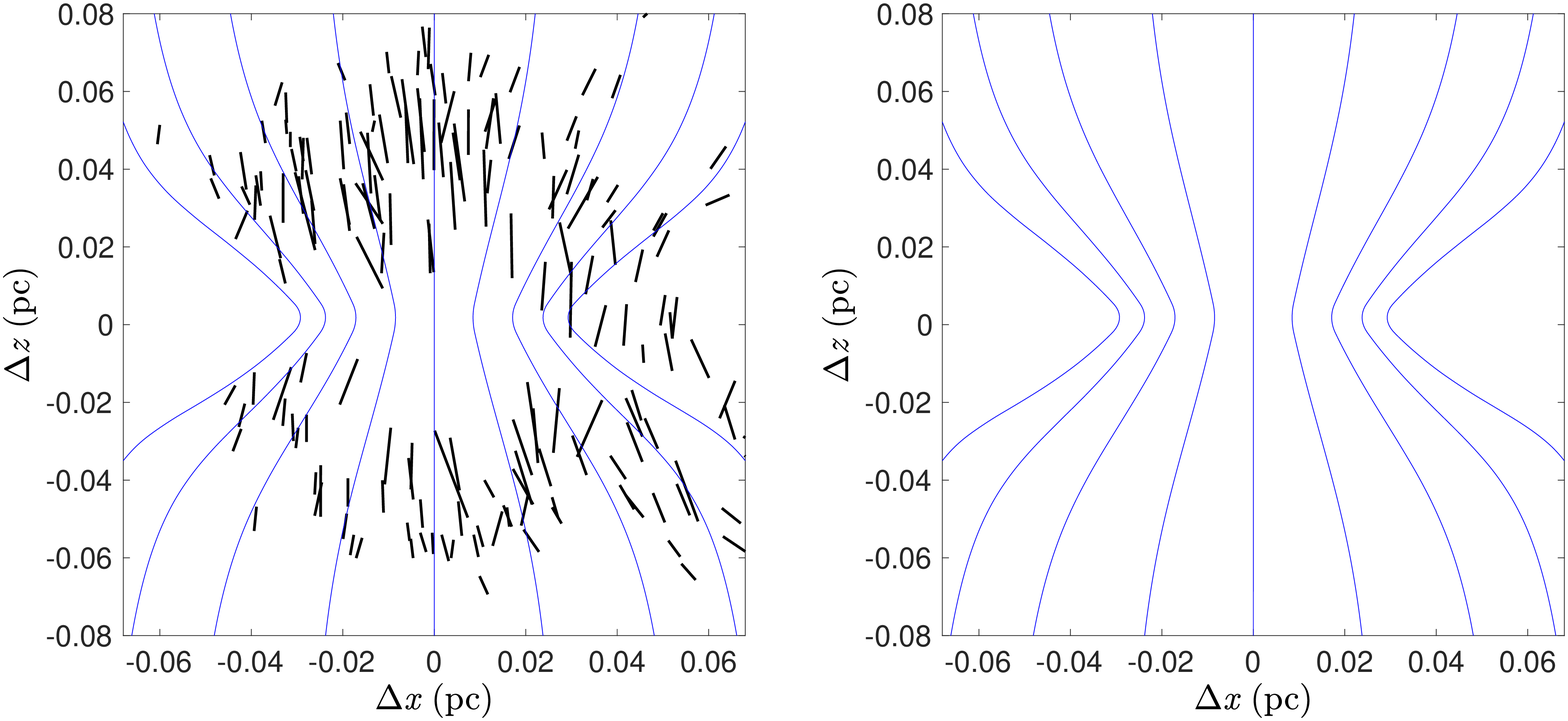}
\end{minipage}
\caption{Model magnetic field fits with a line spacing not indicative of the field strength. Top: Model with $R=0.1$ pc. Bottom: Model with $R=0.2$ pc. Each model is shown in the region where there are observed polarization segments, overlaid with polarization segments on the left, and without polarization segments on the right. }\label{fig:3}
\end{figure*}

\begin{figure*}
 \centering
 \includegraphics[width = \textwidth]{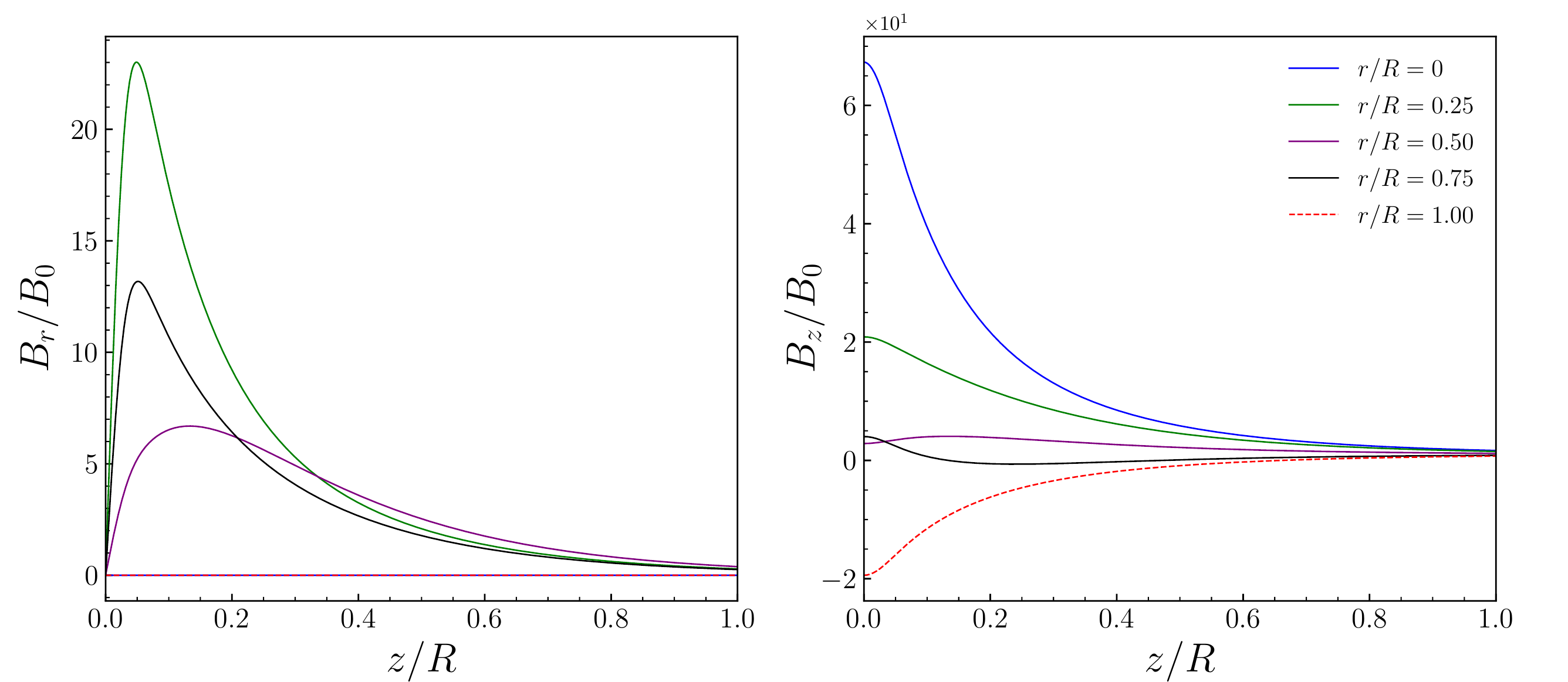}
\end{figure*}
\begin{figure*}
 \centering
 \vspace{-0.4cm}
 \includegraphics[width = 0.98\textwidth]{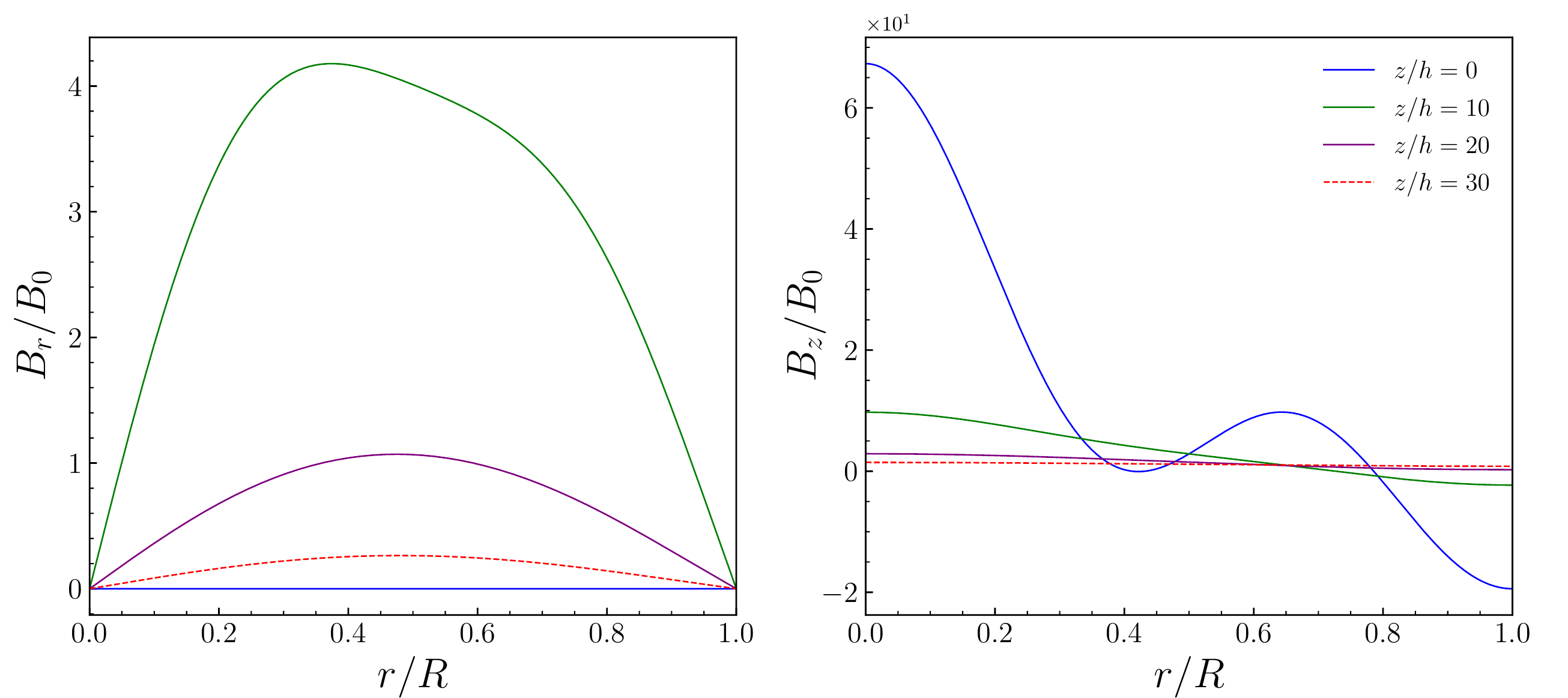}
 \caption{Model magnetic field plot for the model that fixes the core radius at $R = 0.1$ pc, at various slices in the $r$- and $z$-planes. Left: Plots of the $r$-component. Right: Plots of the $z$-component.}
 \label{pos_dep_01}
\end{figure*}

 \begin{figure*}
 \centering
 \includegraphics[width = \textwidth]{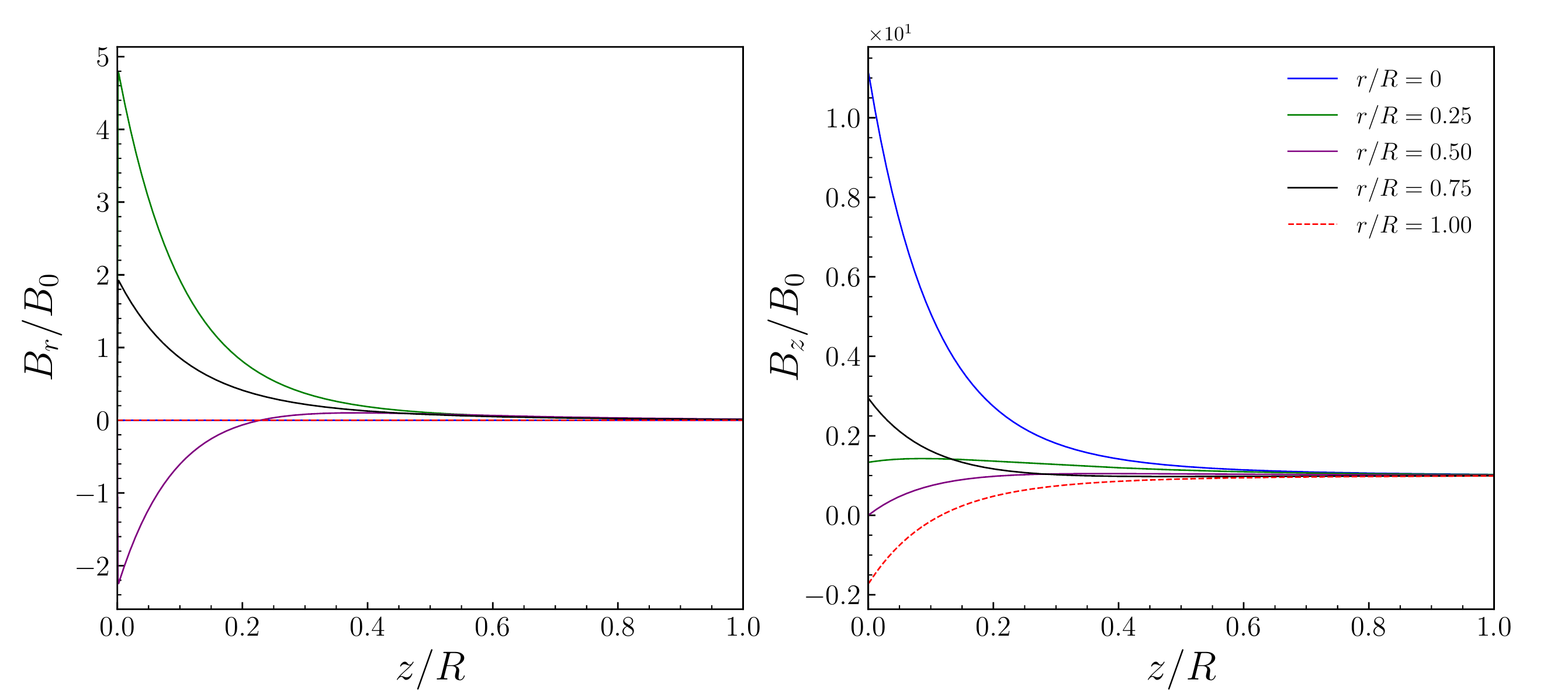}
\end{figure*}
\begin{figure*}
 \centering
 \includegraphics[width = \textwidth]{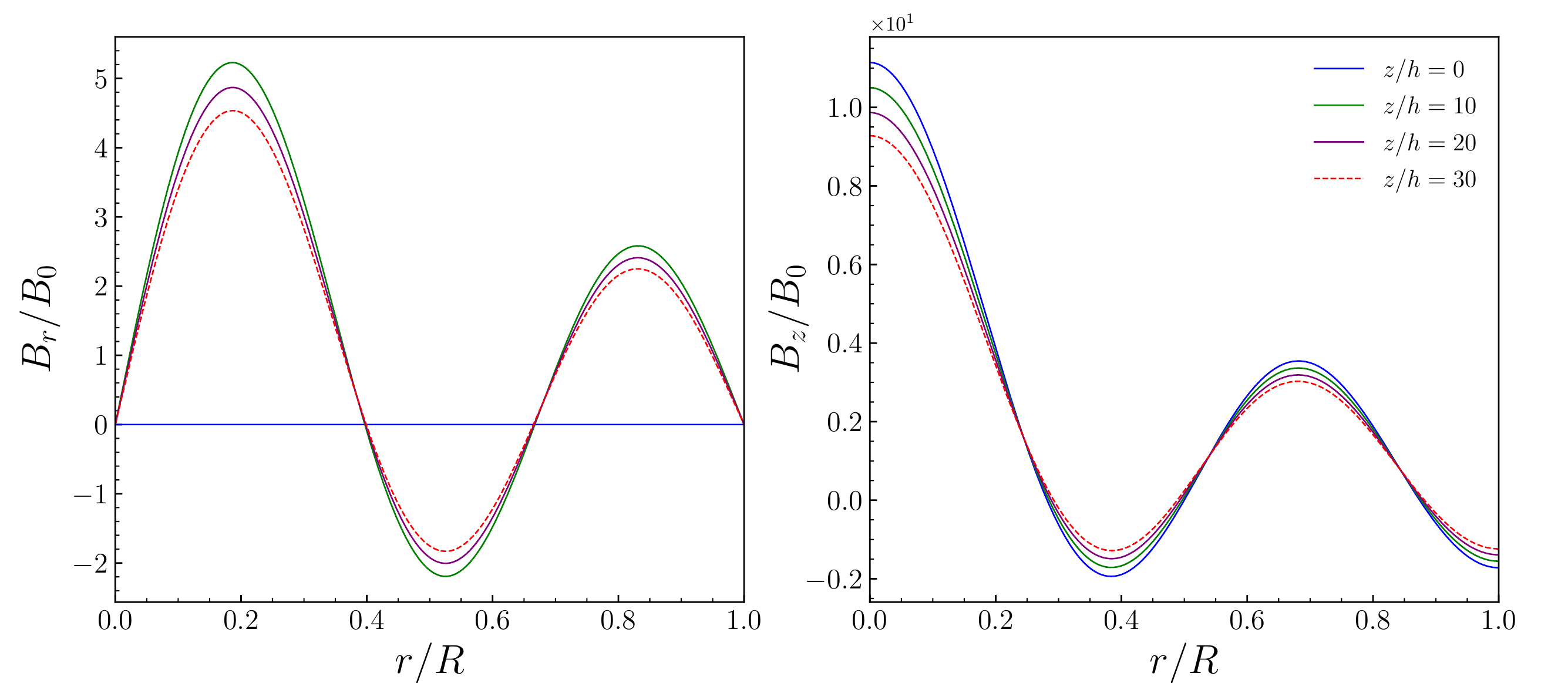}
 \caption{Model magnetic field plot for the model that fixes the core radius at $R = 0.2$ pc, at various slices in the $r$ and $z$-planes. Left: Plots of the $r$-component. Right: Plots of the $z$-component.}
 \label{pos_dep_02}
\end{figure*}
\begin{table}
\begin{tabular}{ |p{4cm}||p{4cm}|  }
 \hline
 \multicolumn{2}{|c|}{Summary of Properties} \\
 \hline
 Observer distance   & 130 pc \\
 Gas mass   & $3.55 \,M_{\odot}$ \\
 Gas temperature &   $9.5$ K\\
 Dust temperature &   $9.5$ K\\
 Grain size & Power-law distribution\\
 Dust composition & Silicate: 62.5 \% \\
 	& Graphite: 37.5 \% \\
 Dust-to-gas mass ratio & 0.01\\
 \hline
\end{tabular}
 \caption{A summary of the general properties of the model used in the POLARIS simulations.} \label{tab:1}
\end{table}

\section{Results}
\label{sec:results}

\subsection{Model Fit}
We use the data set published by \cite{kan17}. Figure \ref{fig:pol} shows the polarimetry data for the core FeSt 1--457, extracted from the data image using Plot digitizer. We observe that the data is not symmetric about $x=0$ and should expect this asymmetry to hinder our model fit in some sense. It is also not exactly symmetric about $z=0$ but this is a lesser discrepancy and we do not concern about it further. Because of the differences in the number and morphology of the polarization segments on the different sides of $x=0$, we perform three different fits of the magnetic field. We fit the entire data set, the left half of the data set and the right half of the data set. The left and right half fits are performed in an effort to see if we can improve the fit by removing the asymmetry in the data, but conclude that both models demonstrate poor fits to modelling the core as a whole. We perform a fit to the entire data set, adopting $R=0.1$ pc, yielding the nondimensionalized parameters
\begin{equation}
\boldsymbol\beta = 
\begin{bmatrix}
15.358 \\
0.502\\
23.464
\end{bmatrix}, \:\:\:\:\:\:\:\: \eta = 0.0366.
\end{equation}
We also perform a fit by fixing the core radius to $R = 0.2$ pc, corresponding physically to a condensation of the core from a significantly larger volume than implied by its density structure.
In this case the optimal dimensionless parameters turn out to be
\begin{equation}
\boldsymbol\beta = 
\begin{bmatrix}
0.611 \\
0.000 \\
4.480
\end{bmatrix} ,\:\:\:\:\:\:\:\: \eta = 7.360 \times 10^{-4}.
\end{equation}
The resulting field lines for both models are shown in Figure \ref{fig:3}. The model’s boundary conditions require 
that $B_r$ will approach zero for $r=R$, i.e., the field becomes vertical at this radial distance. So, physically the model with $R=0.2$ pc allows the curved magnetic field lines to span a larger region than is associated with the observed column density ($\Rgas \approx 0.1$ pc), while the $R=0.1$ pc model fits the curved magnetic field lines in the same region as $\Rgas$. Figures \ref{pos_dep_01} and \ref{pos_dep_02} show the values of $B_r$ and $B_z$ for both models along various cuts along the $r$- and $z$-directions. It can be seen that $B_r$ approaches zero at both $r=R$ and at large heights above the midplane.
The strong degree of curvature in the field lines for offsets $x \gtrsim 0.4$ pc on the right side plane (see Figure \ref{fig:pol}) suggests a weak relative strength of the background magnetic field $B_0$ compared to the core-generated local field. In fact, the dominance of the self-induced magnetic field creates a region of closed magnetic field lines with negative $B_z$ near the midplane and in the region $0.05 \lesssim r \lesssim 0.1$ pc in both model fits.
Figure \ref{mag_comp} demonstrates the magnetic field lines for both model fits overlaid with normalized magnetic field strength ($\sqrt{B_r^2+B_z^2}/B_0$) contours. These plots indicate that the model with smaller radius $R=0.1 $ pc requires a greater relative central concentration of magnetic field strength in order to fit the curvature of the observed field lines.
Figure \ref{fig:series} demonstrates the decomposition of the magnetic field waveform into the three fitted modes. We truncate the series at $m = 3$ as done by \cite{ewe13}. We did make fits with higher order terms, and observed a similar lack of improvement to the fit and in some cases, a reduction in the model performance. The plots show that the $m=1$ mode (half-wavelength) needs to have a significant amplitude in order to fit the overall decline of $B_z$ from center to edge. The high curvature of field lines within the core region emphasizes the $m=3$ mode with little need for much contribution from the $m=2$ mode. 
\subsection{Goodness of Fit}
\label{sec:gof}
\begin{figure*}
 \centering
 \includegraphics[width = \textwidth]{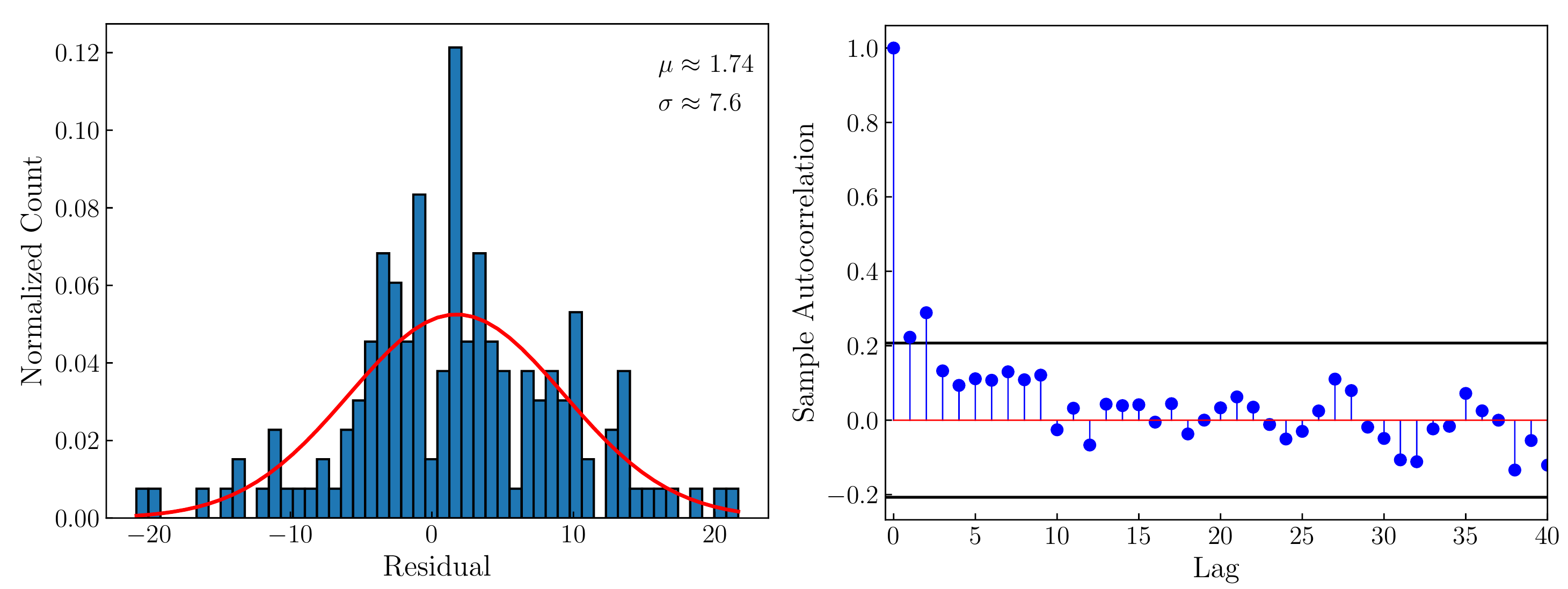}
  \caption{Statistical figures assessing the model's goodness of fit. Left: Histogram of residuals fit to a normal distribution with mean $\mu \approx 1.74$ and standard deviation $\sigma \approx 7.6$. Right: Residual autocorellation function up to 40 lags with 99\% confidence interval.} 
	\label{fig:gof}
 \end{figure*}

\begin{figure*}[h]
 \centering
 \begin{minipage}{1.0\textwidth}
   \includegraphics[width = \textwidth]{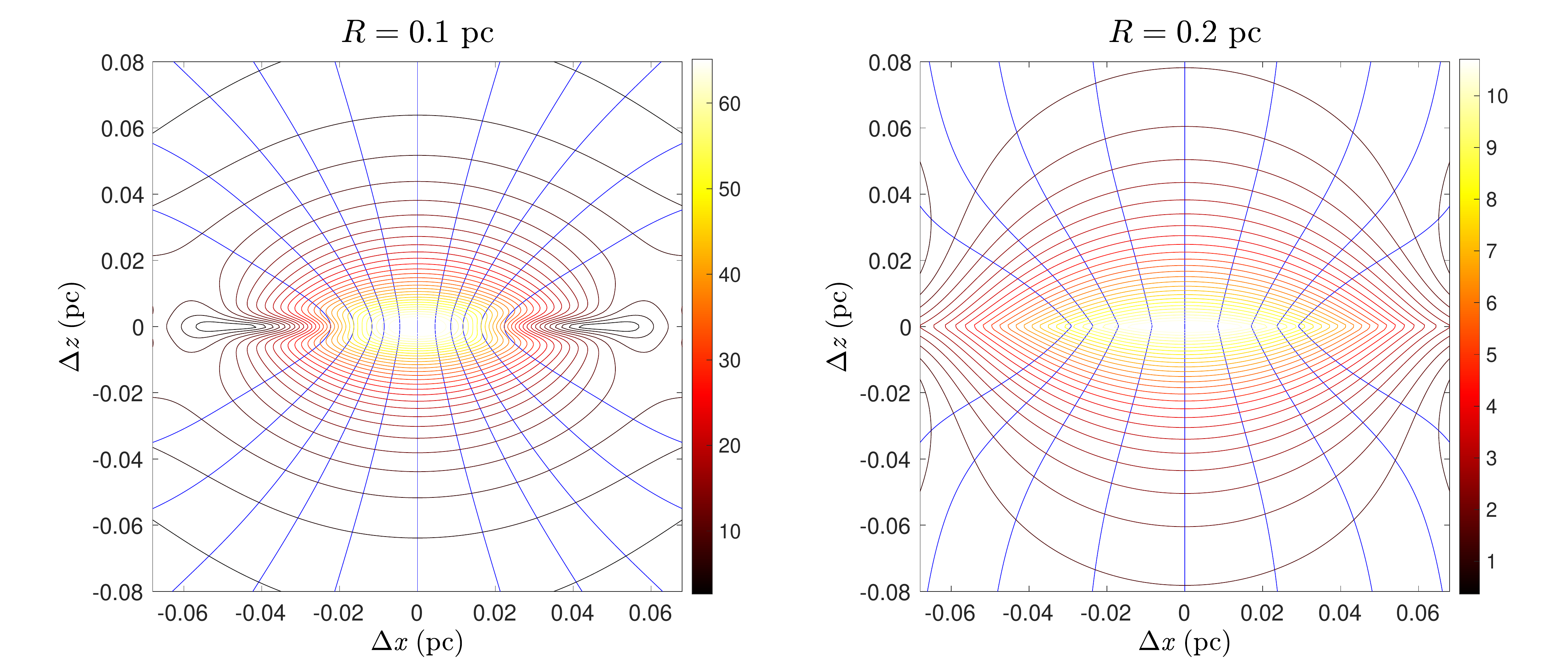}
 \caption{Magnetic field lines overlaid with the normalized magnetic field strength contours. Left: Magnetic field lines for the $R=0.1$ pc model shown in Figure \ref{fig:3}. Right: Magnetic field lines and field strength contours for the model with $R = 0.2$ pc.}  \label{mag_comp}
 \end{minipage}
\vfill
\begin{minipage}{1.0\textwidth}
  \includegraphics[width = \textwidth]{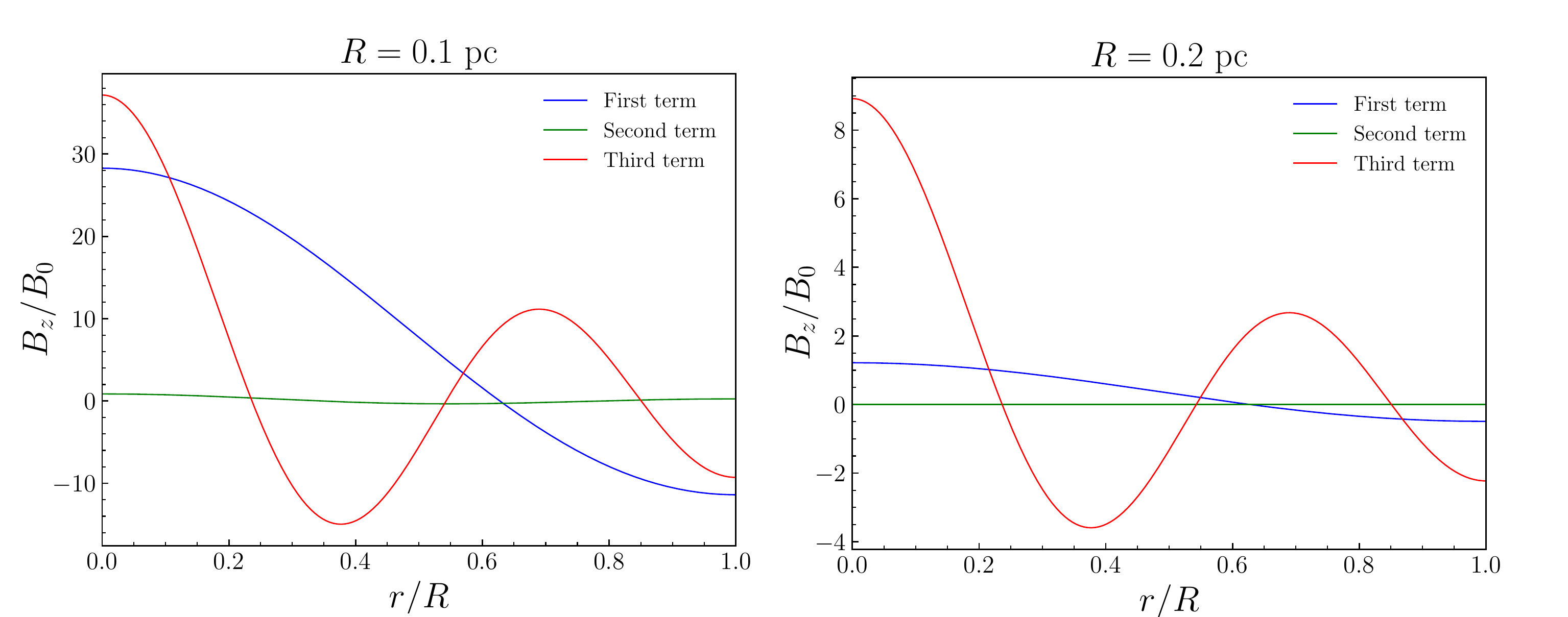}
 \caption{The midplane vertical magnetic field $B_z(r,z=0)$ from each of the first three terms in the series solution. Left: full plane fit for $R=0.1$ pc. Right: full plane fit for the $R = 0.2$ pc model. Each successive term represents a higher frequency term in the Bessel series.} \label{fig:series}
\end{minipage}
\end{figure*}

To assess the goodness of fit, we need to make the standard assumption that the model residuals are normally distributed. In order to test this hypothesis, we perform a $\chi^2$ goodness-of-fit test and a residual whiteness test. Pearson's $\chi^2$-test defines the statistic
\begin{align}
    \chi^2 = \sum_i \frac{(O_i - E_i)^2}{E_i},
\end{align}
where $O_i$ and $E_i$ are representative of the observed and expected counts respectively. After removing the outliers\footnote{Outliers are removed in an effort to compensate for any potential measurement errors or systematic errors in the data extraction process.}
 (we utilize Matlab's \texttt{rmoutliers} function), we achieve a $\chi^2$ statistic of 6.51. This means that the test does not reject the null hypothesis that the model residuals are normally distributed. A histogram of our residuals are demonstrated in the left panel of Figure \ref{fig:gof}. 
 We additionally perform a residual whiteness test using the residual autocorrelation function. If the autocorrelations between residuals at lag $k$ are within a defined confidence interval, then the residuals are uncorrelated. 
 Adopting a 99\% confidence interval (2.58 Gaussian standard deviations) we find the majority of lags ($\sim 95$\%) fall within the interval. Results are shown in the right panel of Figure \ref{fig:gof}. One may ignore the lag at $k = 0$, which is representative of the autocorrelation of a data point with itself.
\subsection{Synthetic Polarization Maps}
\begin{figure*}
 \centering
 \includegraphics[width = \textwidth]{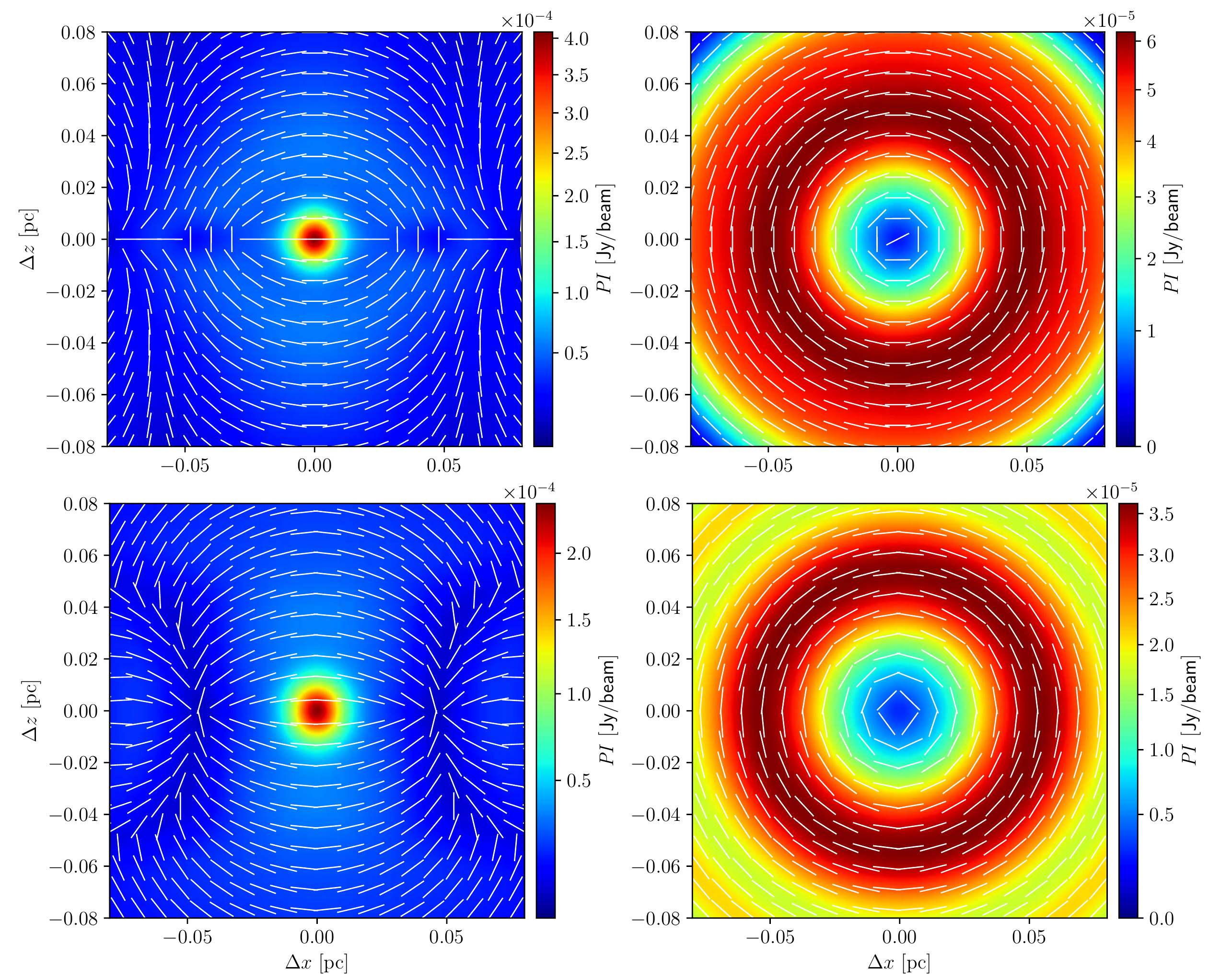}
  \caption{Synthetic polarization maps with electric field vectors in white. The color bar represents the intensity of polarization in Jansky per beam. Left: Oriented with magnetic axis in the plane of the sky, $\theta = 0$. Right: Oriented with magnetic axis along the line of sight, $\theta = \pi/2$. Top: Model with $R=0.1$ pc. Bottom: Model with $R = 0.2$ pc. } \label{pol:f}
 \end{figure*}

POLARIS solves a three dimensional radiative transfer equation for all Stokes parameters:
\begin{equation}
    \frac{d}{dl} \textbf{S} = -\widehat{\textbf{R}}(\alpha)\widehat{\textbf{K}}\widehat{\textbf{R}}(\alpha)^{-1}\textbf{S}+\textbf{J}\, ,
\end{equation}
where $\widehat{\textbf{R}}(\alpha)$ is the rotation matrix, $\widehat{\textbf{K}}$ is the Muller matrix describing the extinction, absorption and scattering, respectively and $\textbf{J}$ is the energy transfer contribution due to emission. The vector $\textbf{S} = \left[ I, Q, U, V \right]^T$ is the Stokes vector having components $I$ and $V$ representing the total intensity and circular polarization, respectively, and the components $Q$ and $U$ representing the states of linear polarization (for further details on the pipeline in POLARIS, refer to the \href{https://www1.astrophysik.uni-kiel.de/~polaris/content/manual.pdf}{software manual}). We run simulations outputting the intensity of total polarization oriented at $\theta = 0$ (magnetic axis in the plane of the sky) and at $\theta = \pi / 2$ (magnetic axis along the line of sight). We simulate both the $R=0.1$ pc and $R = 0.2$ pc models and the results are given in Figure \ref{pol:f}. Additionally, Figure \ref{fig:vecdust} shows the polarization segments rotated by 90 degrees, representative of the inferred magnetic field directions. 
In our simulations, we obtain polarizations as high as 20\% whereas observed polarizations are typically in the range of 4-6 \%. This discrepancy is expected since we have enforced perfect alignment of the dust grains, resulting in a more efficient polarization. Had we adopted an imperfect alignment mechanism, we would expect to see lower polarization, but our purpose here is to study the magnetic field morphology. In that case, the perfect alignment assumption provides the most straightforward path. 

\subsection{Mass-to-Flux Ratio}
We now use the model to study the stability properties of the core. To compute the normalized mass-to-flux ratio 
\begin{equation}
\mu \equiv \frac{(M/ \Phi)}{(M/ \Phi)_{\text{crit}}} = 2\pi \sqrt{G}\left( \frac{M}{\Phi}\right),
\end{equation}
the core's mass and flux must be estimated explicitly. The mass-to-flux ratio gives insight into the contraction mechanisms governing the core formation. 
We calculate the magnetic flux threading the core, 
which we can evaluate efficiently in the equatorial plane ($z=0$) as 
\begin{equation}
\Phi = \int_{S} B_z(r,0) \hat{\textit{\textbf{z}}} \cdot d\textbf{\textit{S}}\, .
\end{equation}
Inserting the dimensionless expression introduced for the model magnetic field, we define the dimensionless flux counterpart to be
\begin{equation}
\widehat{\Phi} = 2\pi \int_0^{\tilde{r}} \tilde{B}_z(\tilde{r}',0) \; \tilde{r}' d \tilde{r}'.
\end{equation}
Inserting Equation \eqref{dim:bz} evaluated at $\tilde{z} =0$ we find
\begin{align}
\widehat{\Phi} = 2\pi \int_0^{\tilde{r}} \left(  \sum_{m=1}^{3} \beta_m J_0(a_{\text{m},1} \tilde{r}')\left[ \text{erfc} \left( \frac{a_{\text{m},1} \eta}{2} \right) \right.  \right. \nonumber \\ \left. \left. + \text{erfc} \left( \frac{a_{\text{m},1} \eta}{2}  \right) \right] + 1 \right)  \; \tilde{r}' d \tilde{r}',
 \label{mf_integral}
\end{align}
which has a closed form solution
\begin{align}
\widehat{\Phi} = \pi \tilde{r}^2\left(\frac{4}{\tilde{r}} \sum_{m=1}^3 \frac{\beta_m}{a_{\text{m},1}} J_1 (a_{\text{m},1}\tilde{r})\, \text{erfc} \left( \frac{a_{\text{m},1} \eta}{2} \right) + 1 \right) .
\end{align}
Inserting $\tilde{r} = 1$ yields the total flux
\begin{align}
    \widehat{\Phi}_{\text {core}} = \pi \left(4 \sum_{m=1}^3 \frac{\beta_m}{a_{\text{m},1}} J_1 (a_{\text{m},1})\, \text{erfc} \left( \frac{a_{\text{m},1} \eta}{2} \right) + 1 \right) .
\end{align}
Since the $a_{\text{m},1}$ are roots of $J_1$, this evaluates to 
$\widehat{\Phi}_{\text{core}} = \pi$,
hence
\begin{equation}
\Phi_{\text{core}} = \pi R ^2 B_0.
\end{equation}
This result shows that our model is fitting the magnetic field structure within $r=R$ and assuming that the cloud condensed from a uniform background field $B_0$ with the fixed magnetic flux shown above. We note that the model allows the field lines to become parallel to the $z$-axis at $r=R$, i.e., $B_r$ goes to zero at $r=R$, however $B_z$ may not converge to $B_0$ at $r=R$ for $z$-values near the midplane \citep[see][\S\ 4 for further discussion]{ewe13}.
The normalized mass-to-flux ratio of the model with $R=0.1$ pc can be written as 
\begin{equation}
\mu = 0.8\, \left( \frac{M_{\rm core}}{3.6\,M_\odot} \right) \left( \frac{R}{0.1\,{\rm pc}} \right)^{-2} \left( \frac{B_0}{15 \, \muG} \right)^{-1} \, .
\end{equation}
For the adopted core mass and radius this means that the core is somewhat subcritical ($\mu_0 \lesssim 1$) if we use the estimated $15\, \muG$ magnetic field strength for the intercore medium of molecular clouds using the OH Zeeman effect \citep{tho19}. This field strength is also similar to an estimate made by \cite{kan18c} using the DCF method and extrapolating a $B-\rho$ relation to the edge of FeSt 1--457. If alternatively we take $B_0 = 5\, \muG$, representative of the general interstellar medium \citep{fer15}, as an estimate for the background field, then the core is somewhat supercritical ($\mu_0 \approx 2$). Given that the mass $M_{\rm core}$ has an estimated $\approx 20\%$ uncertainty \citep{kan05}, the most general conclusion is that the core is transcritical ($\mu_0 \approx 1$) within a factor of about 2. 
We note that an overall value $\mu = 1.41 \pm 0.38$ was estimated by \cite{kan18c} using the DCF method, by measuring deviations of the polarization segments from a parabolic fit to the magnetic field lines. 

 Next, we evaluate Equation (\ref{mf_integral}) for the $R=0.2$ pc model, and use $\tilde{r} = 0.5$ to evaluate the flux within the observed dense core. 
 This leads to $\widehat{\Phi} = 0.44$ for the region $r \leq 0.5 R = 0.1$ pc. Therefore, the normalized mass-to-flux ratio within the half-radius (0.1 pc) is 
\begin{equation}
\mu = 4.6\, \left( \frac{M_{\rm core}}{3.6\,M_\odot} \right) \left( \frac{R}{0.2\,{\rm pc}} \right)^{-2} \left( \frac{B_0}{15 \, \muG} \right)^{-1} \, .
\end{equation}
The model fit with $R=0.2$ pc yields a more decidedly supercritical core. The estimated value of $\mu$ is high enough to imply a supercritical core even accounting for a moderate inclination angle of the magnetic axis relative to the plane of the sky and the uncertainty in the core mass. Furthermore, the value of $\mu$ would be mildly supercritical even if the background magnetic field strength at 0.2 pc is as low as the $5\, \muG$ value \citep{fer15} of the general interstellar medium. The magnetic field of the outer region (0.1 pc $< r < $ 0.2 pc)  of this model fit may imply a subcritical mass-to-flux ratio, however the mass in that region is not well constrained observationally. The model we use is free to fit the magnetic field morphology without making any a priori assumptions about the spatial distribution of the mass-to-flux ratio. 
\subsection{Comparison with Polarization Map}
\begin{figure*}
 \centering
 \includegraphics[width = \textwidth]{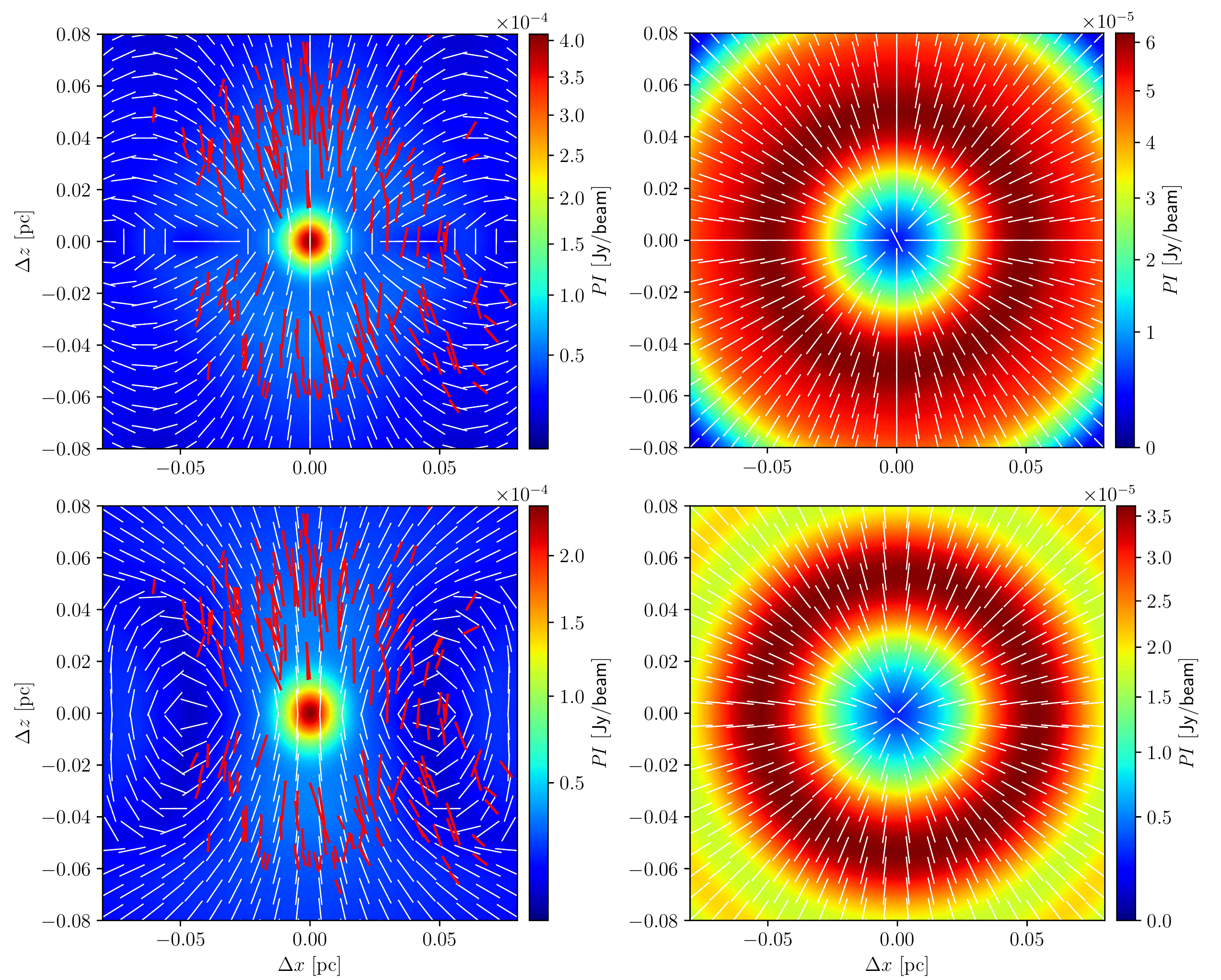}
  \caption{Synthetic polarization maps with magnetic field vectors in white. We orient the polarization segments by 90 degrees to demonstrate the implied morphology of the integrated magnetic field through the core. Left: Oriented with magnetic axis in the plane of the sky, $\theta = 0$. Right: Oriented with magnetic axis along the line of sight, $\theta = \pi/2$. Top: Model with $R=0.1$ pc. Bottom: Model with $R = 0.2$ pc. We additionally demonstrate the superposition of the polarimetry data (red lines) with the inferred magnetic field} \label{fig:vecdust}
 \end{figure*}

The POLARIS simulations yield the degree of polarization as well as its orientation at different locations on the map. These emerge from the full three-dimensional magnetic field structure of our model as well as the adopted density structure and various assumptions about the dust properties listed in Section \ref{sec:polarismodel} and Table \ref{tab:1}.
Since the output from POLARIS is a grid of directions for the polarization of dust emission, we identify this with the direction of the elongation of the dust grains in subsequent discussion. On the other hand the polarization segments due to dichroic extinction measured by \cite{kan17} are perpendicular to the grain elongation. 
The left hand column of Figure \ref{fig:vecdust} shows a comparison of the polarimetry data with the inferred magnetic field from the POLARIS simulation, which are expected to be parallel to one another. This provides a consistency check of whether the three-dimensional magnetic field model that is obtained from fitting the midplane field to the observed polarization segments can also yield a synthetic polarization map that would match the observations. There is a general tendency for the red polarization segments to be close to parallel to the inferred magnetic field direction that is found by rotating the dust grain directions by 90 degrees.
We further perform a residual estimate of the relative angles made with the $z$-axis for the magnetic field directions inferred from the POLARIS output as well as from the best-fit midplane magnetic field model.
Because there are more dust grain vectors in the POLARIS output than polarization segments in the \cite{kan17} dataset, we perform the calculations using only the closest dust grain vector to each polarization segment. For the polarization segments, we rotate the vectors by 90 degrees to get the inferred magnetic field direction. We calculate the difference each set of vectors make with the $z$-axis. For the midplane magnetic field model (Equations (\ref{dim:br}) and (\ref{dim:bz})), we estimate the angle using the relation
$$\varphi = \tan^{-1} \frac{B_r}{B_z},$$
evaluated at the midpoints of the observed polarimetry data. We find that the mean residual angle of the dust grain directions from the synthetic POLARIS map to be $\overline{\varphi} = 0.4081\: \text{rad}$ and the mean residual angle using the model magnetic field to be $\overline{\varphi} = 0.4064\: \text{rad}$.
The good agreement gives us confidence that we can find good fits to the observed polarization segments using a midplane fit of the magnetic model.
\section{Discussion}
\label{sec:discussion}
The observed polarimetry map of the prestellar core FeSt 1--457 reveals a significant curvature of the field lines in the right hand side, which also contains more observed polarization segments than the left side. This is qualitatively consistent with a contraction of the field lines from a significantly larger initial radius than is contained in the map, as was noted by \cite{kan17}. However, when \cite{kan6} fit the polarimetry map with the flux-freezing model of \cite{mye18}, they concluded that the contraction of the core had started at a relatively high density and therefore from a scale much smaller than that associated with the interclump medium of the Pipe Nebula. How can these two conclusions be reconciled?

The flux-freezing model of \cite{mye18} applied to FeSt 1--457 provided a good fit to an inner region of radius $\approx \Rgas/2$ as can be seen from figure 10 of \cite{kan6}. Outside this region, the flux-freezing model reverts to nearly straight parallel field lines and misses the inclined polarization segments seen at $r \gtrsim 0.04$ pc in \autoref{fig:pol}. A parabolic field line fit utilized by \cite{kan17} however does better align with those outer region vectors, since the parabolas become more curved toward the horizontal as the radius increases. So, one interpretation of these different fits is that an approximately flux-freezing contraction started at a relatively dense state and inner length scale, potentially even smaller than the $\Rgas \approx 0.1$ pc adopted from the gas column density profile. The outer highly curved magnetic field lines could also correspond to a separate earlier phase of contraction with approximate flux-freezing. How would the two phases be connected?

Our model fits the magnetic field profile only, with the gas density profile free to follow a different distribution. Therefore, flux freezing from an earlier reference state is not required. Our model with with $R = \Rgas = 0.1$ pc fits the polarization segments, including the outer ones, relatively well, but results in a subcritical mass-to-flux ratio if $B=15\, \muG$ at the core edge, as adopted by \cite{kan18c}. If instead we choose $R = 0.2$ pc, then the relevant mass-to-flux ratio is only that within 0.1 pc, where the density is high and follows a profile similar to a Bonnor-Ebert sphere. Adopting $B=15\, \muG$ for the intercore medium, we find that the inner region is supercritical with $\mu = 4.6$. Outside 0.1 pc the mass-to-flux ratio can have a different value, likely subcritical. The joining of the two different regimes of $\mu$ (inner supercritical and outer subcritical) can be accomplished by ambipolar diffusion, likely a rapid ambipolar diffusion of the kind modeled using large-scale nonlinear flows \citep{li04,nak05,kud08,bas09b,kud11,kud14,aud18,aud19b}. In that scenario a nonlinear flow can bring together ambient subcritical interstellar gas, creating a significant field line curvature, as seen in the outer part of FeSt 1--457. This nonlinear flow need not be symmetric and can possibly contribute to the asymmetry in the observed polarization map. The asymmetry in the data may also be partially or entirely due to the lesser amount of measurable polarization vectors in the outer portion of the left side, due to the nonuniform distribution of background stars.
The innermost dense region undergoes rapid ambipolar diffusion due to the low ionization and strong magnetic field gradient in that region. This initial structure can be filamentary in nature, as also noted by \cite{kan6}. The inner structure, the dense core, becomes supercritical and begins a collapse toward star formation. By fitting our magnetic field model on a scale that is independent of the density structure of FeSt 1--457, we do not impose flux freezing on the overall evolution and can capture the large curvature of the outer region. 

In our fitting of the data, we have not included the possibility of a tilt angle of the core axis with the plane of the sky. This can be incorporated in future modeling where a direct fitting is done between emergent synthetic polarization maps and the observed maps. 
This would allow a better inference of the fully three-dimensional structure characterizing the magnetic field. We could achieve this through a more elaborate fitting protocol using Bayesian statistical inference. This involves computing a posterior distribution of values for each fitting parameter, typically found through a Markov Chain Monte Carlo technique. While computationally intensive, this methodology benefits from employing a grid of models, all of which can be processed through POLARIS. The input models could be three-dimensional simulations as well as the analytic model presented in this paper. 


\section{Conclusions}
\label{sec:conclusion}
We have compared a polarimetry map of the prestellar core FeSt 1--457 with a single plane of a model magnetic field as well as with an integrated synthetic map of polarization directions. There is a very good agreement between the magnetic field taken from the midplane of the model with the inferred directions that emerge from a synthetic dust emission polarization map obtained using the POLARIS code. 
Our model fits conclude that FeSt 1--457 is in a transcritical or mildly supercritical state. The strong curvature of field lines seen in the right side of the core imply that the magnetic field profile has condensed from a large distance from the current core size as defined by the density distribution. By adopting a scale length $R$ of the magnetic field that is twice $\Rgas$, we can fit the highly curved field lines while finding a supercritical mass-to-flux ratio within the region $r \leq \Rgas$. The region $r > \Rgas$ may be subcritical, and the transition in mass-to-flux ratio was likely accomplished by rapid ambipolar diffusion that followed a core assembly by a large-scale flow. 

This study demonstrates the capabilities and uses of an analytic model like the one developed by \citet{ewe13}. Fitting the direction of the polarization segments by either the ratio $B_r / B_z$  evaluated in the model midplane or with an integrated synthetic polarization map holds the promise of inferring the magnetic field strength in units of a background field $B_0$. If $B_0$ is estimated separately, then this method calculates the magnetic field strength at all interior points, in a manner completely independent of the DCF method. In fact the DCF method can be used in the future in conjunction with analytic model fits in order to better estimate the magnetic field fluctuations. This model utilizes a physically plausible vertical Gaussian distribution of electric current density, however future work can also explore more general distributions.

\acknowledgments
We thank the anonymous referee for comments that improved the manuscript. We also thank Sayantan Auddy and Indrani Das for comments. SB is supported by a Discovery Grant from NSERC.
\bibliography{manuscript}{}
\bibliographystyle{aasjournal}
\end{document}